\documentclass[aps,prl,twocolumn,superscriptaddress,floatfix,a4paper]{revtex4}
\usepackage{amssymb}
\usepackage{amsfonts}

\usepackage{graphicx,graphics,epsfig}
\usepackage{dcolumn}
\usepackage{bm}
\usepackage{amsmath}
\usepackage{verbatim}
\usepackage{color}
\usepackage{subfigure}
\usepackage{times,natbib}
\usepackage{amsmath,amsfonts,amssymb,graphics,graphicx,epsfig,color,times,natbib}

\begin{document}

\title{Quantum No-Cloning Theorem Certified by Bell's Theorem}

\author{Jing-Ling Chen}
 \email{chenjl@nankai.edu.cn}
\affiliation{Theoretical Physics Division, Chern Institute of
Mathematics, Nankai University, Tianjin 300071, People's Republic of
China} \affiliation{Centre for Quantum Technologies, National
University of Singapore, 3 Science Drive 2, Singapore 117543}

\author{Hong-Yi Su}
\affiliation{Theoretical Physics Division, Chern Institute of
Mathematics, Nankai University, Tianjin 300071, People's Republic of
China} \affiliation{Centre for Quantum Technologies, National
University of Singapore, 3 Science Drive 2, Singapore 117543}

\author{Chunfeng Wu}
\affiliation{Centre for Quantum Technologies, National
University of Singapore, 3 Science Drive 2, Singapore 117543}

\author{Xiang-Jun Ye}
\affiliation{Theoretical Physics Division, Chern Institute of
Mathematics, Nankai University, Tianjin 300071, People's Republic of
China}

\author{Dong-Ling Deng}
\affiliation{Department of Physics and MCTP, University of Michigan,
Ann Arbor, Michigan 48109, USA}

\author{C. H. Oh}
 \email{phyohch@nus.edu.sg}
\affiliation{Centre for Quantum Technologies, National University of
Singapore, 3 Science Drive 2, Singapore 117543}
\affiliation{Department of Physics, National University of
Singapore, 2 Science Drive 3, Singapore 117542}

\date{\today}

\begin{abstract}
We investigate the connection between quantum no-cloning theorem and
Bell's theorem. Designing some Bell's inequalities, we show that
quantum no-cloning theorem can always be certified by Bell's
theorem, and this fact in turn reflects that our physical world is
essentially nonlocal.
\end{abstract}


\maketitle


In classical information theory, one may duplicate an unknown state
into many copies, by measuring these copies one may accurately
reveal the information of this unknown state. However, the situation
dramatically changes when one turns to the quantum domain. Due to
the Copenhagen probability interpretation of the
quantum-state-collapse, contrary to the classical case, it is
impossible to determine with certainty an unknown quantum state when
one owns only one copy of it. When measuring a quantum state,
state-collapse happens so that one gets different measurement values
of the eigenstates with probabilities. To experimentally identify a
quantum state, one needs the process of quantum state tomography:
one has to measure the states prepared by the same device for large
numbers of times so as to get the information of probabilities
corresponding to each eigenstate. Moreover, the situation becomes
even more worse because one cannot indeed make a perfect copy of an
unknown quantum state, as indicated by the \emph{quantum no-cloning
theorem} \cite{Woot1982, gisinRMP, Buzek1996}.

Quantum no-cloning theorem can be proved by using the superposition
principle and the linearity of quantum transformations. The
impossibility of perfect cloning seems at first annoying
restriction, but it can be used favourably. During the nearly 30
years since its discovery, the no-cloning theorem has had a
significant impact on the development of quantum information theory:
it renders the classical error corrections futile on quantum states
\cite{Shor, Steane}, and plays a vital role in the security of
quantum cryptography \cite{BB84}.
On the other hand, Bell's theorem has been regarded as ``the most
profound discovery in science" \cite{Stapp}. Based on local hidden
variable model, multipartite systems separated by arbitrarily long
distance satisfy some classical bounds in the form of Bell's
inequalities, and the violation of which thus implies quantum
nonlocality \cite{Bell, Horodecki09}.
As basic tenets in quantum information theory, the no-cloning
theorem and Bell's theorem have profound implications in quantum
information and related fields. Despite these significant impacts,
their deeper implication however has yet to be explored. In this
work we investigate the connection between quantum no-cloning
theorem and Bell's theorem. With the presence of some Bell's
inequalities, we show that quantum no-cloning theorem can always be
certified by Bell's theorem.

Let us briefly
review the quantum no-cloning theorem. In the simplest case of one
qubit, initially one has as input an arbitrary Schrodinger's cat
state
\begin{eqnarray}
|\psi^{\rm sc}\rangle=\cos\frac{\xi}{2}\;|0\rangle+\sin\frac{\xi}{2}
e^{i\varphi}\;|1\rangle,
\end{eqnarray}
where the parameters $\xi\in[0,\pi]$, $\varphi\in[0,2\pi]$, and
$|0\rangle$ and $|1\rangle$ are orthonormal bases for the
two-dimensional system. If one has an operation $\mathcal {U}_2$
duplicating the states $|0\rangle$ and $|1\rangle$ in the following
way: $\mathcal {U}_2|0\rangle|0\rangle\rightarrow
|0\rangle|0\rangle,\;\;\mathcal {U}_2|1\rangle|0\rangle\rightarrow
|1\rangle|1\rangle$, then this yields an output quantum state as
\begin{eqnarray}
\mathcal {U}_2|\psi^{\rm
sc}\rangle|0\rangle\rightarrow|\psi_2\rangle= \cos\frac{\xi}{2}\;
|0\rangle|0\rangle+\sin\frac{\xi}{2}
e^{i\varphi}\;|1\rangle|1\rangle,\label{2qubitqc}
\end{eqnarray}
which is obviously not equal to $|\psi^{\rm sc}\rangle\otimes
|\psi^{\rm sc}\rangle$. Therefore the operation $\mathcal {U}_2$ is
not a quantum cloning machine (QCM), or a perfect $1\rightarrow 2$
quantum cloning is impossible. It is clear that $|\psi_2\rangle$ is
an entangled pure state of two qubits except $\xi=0, \pi$,
it violates the Clauser-Horne-Shimony-Holt (CHSH) inequality
\cite{CHSH} that reads
\begin{eqnarray}
\frac{1}{2}(Q_{11}+Q_{12}+Q_{21}-Q_{22})\leq 1,\label{CHSH}
\end{eqnarray}
where $Q_{ij}\;(i,j=1,2)$ are correlation functions for two
subsystems. The violation of the CHSH inequality by $|\psi_2\rangle$
tells us the existence of nonlocality. Recall that $|\psi_2\rangle$
is generated by the action of $\mathcal {U}_2$ on one-qubit
Schr{\"o}dinger's cat state, such a nonlocal phenomenon of
$|\psi_2\rangle$ shows the impossibility of perfect quantum cloning.
We then extend the consideration to $N$-qubit case. If there is an
operation $\mathcal {U}_N$ such that $\mathcal
{U}_N|0\rangle|0\rangle\cdots|0\rangle\rightarrow
|0\rangle|0\rangle\cdots|0\rangle,\;\;\mathcal
{U}_N|1\rangle|0\rangle\cdots|0\rangle\rightarrow
|1\rangle|1\rangle\cdots|1\rangle$, acting on the initial
Schrodinger's cat state, we have
\begin{eqnarray}
&&\mathcal {U}_N|\psi^{\rm sc}\rangle|0\rangle\cdots|0\rangle\rightarrow\nonumber\\
&&|\psi_N\rangle= \cos\frac{\xi}{2}\;
|0\rangle|0\rangle\cdots|0\rangle+\sin\frac{\xi}{2}
e^{i\varphi}\;|1\rangle|1\rangle\cdots|1\rangle.\label{Nqubitqc}
\end{eqnarray}
The state $|\psi_N\rangle$ is an $N$-qubit entangled pure state
rather than a separable state $|\psi^{\rm
sc}\rangle\otimes|\psi^{\rm sc}\rangle\otimes\cdots|\psi^{\rm
sc}\rangle$. This tells us that $\mathcal {U}_N$ is not a perfect
$1\rightarrow N$ QCM. According to the well-known Gisin's theorem
\cite{Gisin, CHSH, WWZB, us, Marcin, Fei}, any entangled pure state
always violates some Bell's inequalities, thus the non-existence of
$1\rightarrow N$ perfect QCM can be certified by Bell's theorem.
Then a natural question arises: Can this certification hold for
mixed-state case?

\begin{figure}[tbp]
\includegraphics[width=89mm]{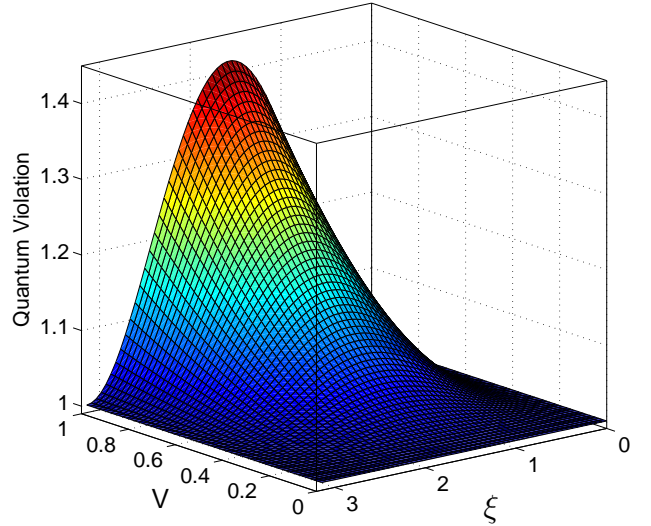}\\
\caption{(Color online) Quantum violations of the CHSH inequality by
$\rho_2$ versus $\xi$ and $V$. Except for the points at
$V=0$, $\xi=0,\;\pi$, the quantum prediction of the left-hand side
of the CHSH inequality (\ref{CHSH}) is always larger than 1. The
point of $V=0$ represents the two-qubit colored noise and the points
of $\xi=0,\;\pi$ reflect separable states.} \label{fig1}
\end{figure}

In practical quantum information and computation protocols, noise is
always an inevitable issue. In the presence of white noise,
initially the input one-qubit Schr{\"o}dinger's cat state is in a
mixed state as follows
\begin{eqnarray}
\rho^{\rm sc}=V\rho_0+\frac{1-V}{2}I,\label{onecopy}
\end{eqnarray}
where $\rho_0=|\psi^{\rm sc}\rangle\langle\psi^{\rm sc}|$,
$I=\sum_{m=0}^1|m\rangle\langle m|$ is one-qubit identity matrix,
$I/2$ represents the density matrix of white noise for single qubit,
and $V\in[0,1]$ the so-called visibility. Furthermore one can recast
$\rho^{\rm sc}$ to a form as $(1+\vec{r}\cdot\vec{\sigma})/2$ and
finds that $\rho^{\rm sc}$ has indeed run over all the one-qubit
states in the Bloch sphere, here $\vec{\sigma}$ is Pauli matrix
vector and $\vec{r}$ is the Bloch vector. Through the action of
$\mathcal {U}_2$ on $\rho^{\rm sc}$, one gets an output state as
\begin{eqnarray}
\rho_2=V|\psi_2\rangle\langle \psi_2|+\frac{1-V}{2}I_2^{\rm cn},
\end{eqnarray}
where $I_2^{\rm cn}/2=(|00\rangle\langle00|+|11\rangle\langle11|)/2$
is the density matrix denoting colored noise for two qubits. It is
obvious that $\rho_2$ is not equal to $\rho^{\rm sc}\otimes\rho^{\rm
sc}$. Let us examine the violation of the CHSH inequality by the
state $\rho_2$. The quantum prediction for correlation function
reads $Q_{ij}={\rm Tr}(\rho_2
\vec{\sigma}_{\hat{n}_{A_i}}\otimes\vec{\sigma}_{\hat{n}_{B_j}})$,
with $\vec{\sigma}_{\hat{n}_{A_i}}=\hat{n}_{A_i}\cdot\vec{\sigma}$,
$\vec{\sigma}_{\hat{n}_{B_j}}=\hat{n}_{B_j}\cdot\vec{\sigma}$
$(i,j=1,2)$,
$\hat{n}_{A_i}=(\sin\theta_{A_i}\cos\phi_{A_i},\sin\theta_{A_i}\sin\phi_{A_i},\cos\theta_{A_i})$
and
$\hat{n}_{B_j}=(\sin\theta_{B_j}\cos\phi_{B_j},\sin\theta_{B_j}\sin\phi_{B_j},\cos\theta_{B_j})$
$(i,j=1,2)$ are unit vectors describing measurement directions for
subsystems $A$ and $B$ respectively. Choose the measurement setting
as
$\phi_{A_1}=\phi_{B_1}=\phi_{B_2}=\theta_{A_1}=0,\;\;\theta_{A_2}=\pi/2,\;\;\theta_{B_2}=-\theta_{B_1},\;\;\phi_{A_2}=\varphi$,
the left-hand side of the CHSH inequality (\ref{CHSH}) becomes
$(\cos\theta_{B_1}+V \sin\xi\sin\theta_{B_1})\leq\sqrt{1 + (V \sin
\xi)^2}$. In Fig. \ref{fig1} we plot the variation of the quantum
violation versus visibility $V$ and parameter $\xi$. It is clear
that, except for the points at $V=0$, $\xi=0,\;\pi$, the output
state $\rho_2$ always violates the CHSH inequality (\ref{CHSH}) and
hence exhibits there is no perfect $1\rightarrow 2$ QCM. This result
is remarkable: it tells us that the non-existence of perfect
$1\rightarrow 2$ QCM can be certified by Bell's theorem even when
initially the one-qubit Schr{\"o}dinger's cat state is in the
presence of white noise.

The above conclusion can be extended to the case of perfect
$1\rightarrow N$ QCM. Consider the state $\rho_N$ produced from
$\rho^{\rm sc}$ by the action of $\mathcal {U}_N$, we have
\begin{eqnarray}
\rho_N=V|\psi_N\rangle\langle \psi_N|+\frac{1-V}{2}I_N^{ \rm cn},
\end{eqnarray}
where 
$I_N^{ \rm
cn}=|0...0\rangle\langle0...0|+|1...1\rangle\langle1...1|$. We have
the following two theorems.

\emph{Theorem 1.}---For even-$N$, the non-existence of $1\rightarrow
N$ perfect QCM can be certified by Bell's inequalities:
\begin{eqnarray}
\mathcal {I}^N=\frac{1}{2^{N-1}}\biggr(\sum_{i_1,i_2,...,i_N=1}^2
Q_{i_1i_2...i_N}\biggr)-Q_{22...2}\leq 1,\label{BI-N}
\end{eqnarray}
where correlation functions $Q_{i_1i_2...i_N}=X_{1i_1}X_{2i_2}\cdots
X_{Ni_N}$, $X_{ni_n}=\pm 1$, and $i_n$ indicating settings for each
subsystems.

\emph{Proof.}---For $N=2$, the inequality (\ref{BI-N}) reduces to
the CHSH inequality, thus we need to prove Theorem 1 from $N=4$. We
firstly consider the four-qubit case and then generalize to
arbitrary even $N$.  For $N=4$, the inequality (\ref{BI-N}) reads
\begin{eqnarray}
\mathcal
{I}^4&=&\frac{1}{8}(Q_{1111}+Q_{1112}+...+Q_{2222})-Q_{2222}\leq1.\label{BI-4}
\end{eqnarray}
Given $Q_{ijkl}=a_ib_jc_kd_l$, terms in the above bracket are simply
the expansion of $\mathcal
{Y}=\frac{1}{8}(a_1+a_2)(b_1+b_2)(c_1+c_2)(d_1+d_2)$, which may take
only three distinct values of $\{-2,0,2\}$ because
$a_i,b_j,c_k,d_l=\pm 1$. Moreover, the value of $Q_{2222}$ is
related to $\mathcal {Y}$, hence the quantity $\mathcal {I}^4$ has a
certain upper bound. For instance, when $\mathcal {Y}=2$, then
$Q_{2222}=a_2b_2c_2d_2=1$, so $\mathcal {I}^4=1$; when $\mathcal
{Y}=-2$, then $Q_{2222}=-1$, so $\mathcal {I}^4=-1$; when $\mathcal
{Y}=0$, then $\mathcal {I}^4\leq1$ since $Q_{2222}$ is either 1 or
$-1$. Therefore the inequality $\mathcal {I}^4\leq1$ holds for any
situation.

Quantum mechanically, the correlation function consists of two
parts: $Q_{ijkl}=V\mathcal {Q}_1+(1-V)\mathcal {Q}_2$ where
$\mathcal {Q}_1={\rm Tr}(|\psi_4\rangle\langle\psi_4|\vec{\sigma}_{\hat{n}_{A_i}}
\otimes\vec{\sigma}_{\hat{n}_{B_j}}\otimes\vec{\sigma}_{\hat{n}_{C_k}}\otimes\vec{\sigma}_{\hat{n}_{D_l}})$,
$\mathcal {Q}_2=\frac{1}{2}{\rm Tr}(I_4^{ \rm
cn}\vec{\sigma}_{\hat{n}_{A_i}}
\otimes\vec{\sigma}_{\hat{n}_{B_j}}\otimes\vec{\sigma}_{\hat{n}_{C_k}}\otimes\vec{\sigma}_{\hat{n}_{D_l}})$.
The relative phase angle $\varphi$ in the state $|\psi_N\rangle$ can be set to zero by some local unitary
transformations, for our purpose.  For simplicity, hereafter the
relative phase is simply omitted. For four-qubit case, calculation
result shows
\begin{eqnarray}
\mathcal {Q}_1&=&\cos\theta_{a_i}\cos\theta_{b_j}\cos\theta_{c_k}\cos\theta_{d_l}+\sin\xi\sin\theta_{a_i}\sin\theta_{b_j}\nonumber\\
&&\;\;\;\;\;\;\times\sin\theta_{c_k}\sin\theta_{d_l}\cos(\phi_{a_i}+\phi_{b_j}+\phi_{c_k}+\phi_{d_l}),\nonumber\\
\mathcal
{Q}_2&=&\cos\theta_{a_i}\cos\theta_{b_j}\cos\theta_{c_k}\cos\theta_{d_l},
\end{eqnarray}
and thus
$Q_{ijkl}=\cos\theta_{a_i}\cos\theta_{b_j}\cos\theta_{c_k}\cos\theta_{d_l}+V\sin\xi\sin\theta_{a_i}\sin\theta_{b_j}\sin\theta_{c_k}\sin\theta_{d_l}$.
In the following, we would like to choose some appropriate settings
to simplify the expression of $\mathcal {I}^4$. These settings are
not the optimal settings that lead to the really maximal violation
of $\mathcal {I}^4$, but  sufficient to show clearly the quantum
violation of the inequality (\ref{BI-4}). We choose the settings as
follows:
$\phi_{a_i}=\phi_{b_j}=\phi_{c_k}=\phi_{d_l}=0,\;\;\theta_{d_i}=\theta_{c_i}=\theta_{b_i}=\pi-\theta_{a_i}$,
$(i, j, k, l=1, 2)$, then
according to (\ref{BI-4}), the quantum mechanical expression of
$\mathcal {I}^4$ reads
\begin{eqnarray}
\mathcal {I}^4_{\rm QM}&=&\frac{1}{8}[-(\cos\theta_{a_1}+\cos\theta_{a_2})^4\nonumber\\
&&+V\sin\xi(\sin\theta_{a_1}+\sin\theta_{a_2})^4\nonumber\\
&&+8\cos^4\theta_{a_2}-8V\sin\xi\sin^4\theta_{a_2}].
\end{eqnarray}
Next, let $\theta_{a_2}=\pi$, we have
\begin{eqnarray}
\mathcal {I}^4_{\rm QM}&=&\frac{1}{8}[8+V\sin\xi\sin^4\theta_{a_1}-(1-\cos\theta_{a_1})^4]\nonumber\\
&=&1+\frac{V\sin\xi\sin^4\theta_{a_1}}{8}\biggr(1-\frac{\tan^4\frac{\theta_{a_1}}{2}}{V\sin\xi}\biggr).
\end{eqnarray}
One can verify that when
$0<\theta_{a_1}<2\arctan[(V\sin\xi)^{1/4}]$, $\mathcal {I}^4_{\rm
QM}$ is always larger than 1. Hence the non-existence of
$1\rightarrow 4$ perfect QCM can be certified by Bell's inequality
$\mathcal {I}^4 \leq 1$.

The generalization to arbitrary even-$N$ case is straightforward.
The summation term in (\ref{BI-N}) is binomial expansion of
$\mathcal
{Y}=\frac{1}{2^{N-1}}(X_{11}+X_{12})(X_{21}+X_{22})\cdots(X_{N1}+X_{N2})$,
which may take only three distinct values of $\{-2,0,2\}$, and the
value of $Q_{22...2}$ is related to $\mathcal {Y}$. Similar to the
analysis of the four-qubit case, when $\mathcal {Y}=2$, then
$Q_{22...2}=1$, so $\mathcal {I}^N=1$; when $\mathcal {Y}=-2$, then
$Q_{22...2}=-1$, so $\mathcal {I}^N=-1$; when $\mathcal {Y}=0$, then
$\mathcal {I}^N\leq1$ since $Q_{22...2}$ is no larger than 1. Hence
inequality $\mathcal {I}^N\leq1$ is always fulfilled in local hidden
variable description. In quantum mechanics, we similarly have
\begin{eqnarray}
Q_{i_1i_2...i_N}&=&\cos\theta_{1i_1}\cos\theta_{2i_2}\cdots\cos\theta_{Ni_N}\nonumber\\
&&+V\sin\xi\sin\theta_{1i_1}\sin\theta_{2i_2}\cdots
\sin\theta_{Ni_N}\nonumber\\
&&\times\cos(\phi_{1i_1}+\phi_{2i_2}+\cdots+\phi_{Ni_N}).
\end{eqnarray}
After setting $\phi_{1i_1}=\phi_{2i_2}=\cdots=\phi_{Ni_N}=0$,
$\theta_{N1}=\cdots=\theta_{21}=\pi-\theta_{11}$,
$\theta_{N2}=\cdots=\theta_{22}=\pi-\theta_{12}$, and
$\theta_{12}=\pi$, we obtain the following quantum expression as
\begin{eqnarray}
\mathcal {I}^N_{\rm QM}&=&\frac{1}{2^{N-1}}[2^{N-1}+V\sin\xi\sin^N\theta_{11}-(1-\cos\theta_{11})^N]\nonumber\\
&=&1+\frac{V\sin\xi\sin^N\theta_{11}}{2^{N-1}}\biggr(1-\frac{\tan^N\frac{\theta_{11}}{2}}{V\sin\xi}\biggr).
\end{eqnarray}
Clearly, when $0<\theta_{11}<2\arctan[(V\sin\xi)^{1/N}]$, $\mathcal
{I}^N_{\rm QM}$ is always larger than 1.
Theorem 1 is henceforth proved.

\emph{Theorem 2.}---For odd-$N$, the non-existence of $1\rightarrow
N$ perfect QCM can be certified by Bell's inequalities:
\begin{eqnarray}
\mathcal
{I}^N=\frac{1}{2^{N-1}}\biggr(\sum_{i_1,...,i_{N-1}=1}^2\sum_{i_{N}=0}^1
Q_{i_1i_2...i_N}\biggr)-Q_{22...20}\leq 1,\nonumber\\\label{BI-N2}
\end{eqnarray}
with $i_N=0$ indicating that no measurement is performed on the
$N$-th qubit.

\emph{Proof.}---Actually, Bell's inequality $\mathcal {I}^N\leq 1$
in (\ref{BI-N}) also holds classically for the odd-$N$ case.
However, it cannot detect all the non-existence of $1\rightarrow N$
perfect QCM for the odd-$N$ case when the visibility $V$ is
sufficiently small, and due to this reason we resort to other form
of Bell's inequalities. Substituting $X_{N2}=1$ into inequality
(\ref{BI-N}), one then arrives at Bell's inequality (\ref{BI-N2}),
which can certify that there are no $1\rightarrow {\rm odd}$-$N$
perfect QCM except $V=0$, $\xi=0,\;\pi$.

Quantum mechanically, the inequality (\ref{BI-N2}) consists of two
kinds of correlation functions, that is, full $N$-particle
correlation $Q={\rm Tr}(\vec{\sigma}_{\hat{n}_1}
\otimes\vec{\sigma}_{\hat{n}_2}\otimes\cdots\otimes\vec{\sigma}_{\hat{n}_N}\rho_N)=
V\cos\xi\cos\theta_{1i_1}\cos\theta_{2i_2}\cdots\cos\theta_{Ni_N}+V\sin\xi\sin\theta_{1i_1}\sin\theta_{2i_2}\cdots
\sin\theta_{Ni_N}\cos(\phi_{1i_1}+\phi_{2i_2}+\cdots+\phi_{Ni_N})$
and $(N-1)$-particle correlation $Q'={\rm
Tr}(\vec{\sigma}_{\hat{n}_1}
\otimes\vec{\sigma}_{\hat{n}_2}\otimes\cdots\otimes\vec{\sigma}_{\hat{n}_{N-1}}\otimes
\textbf{1 }\;\rho_N)
=\cos\theta_{1i_1}\cos\theta_{2i_2}\cdots\cos\theta_{Ni_N}$. After
setting $\phi_{1i_1}=\phi_{2i_2}=\cdots=\phi_{Ni_N}=0$,
$\theta_{ni_n}=\frac{\pi}{2}\times i_n,\;(n\neq1)$, and
$\theta_{12}=0$, then one finds that
$Q_{111...11}=V\sin\xi\sin\theta_{11},\;\;Q_{122...20}=-\cos\theta_{11},\;\;Q_{222...20}=-1$,
and the others are zeros. Hence the quantum expression of $\mathcal
{I}^N$ in (\ref{BI-N2}) becomes
\begin{eqnarray}
\mathcal {I}^N_{\rm QM}
&=&1+\frac{1}{2^{N-1}}(V\sin\xi\sin\theta_{11}-\cos\theta_{11}-1)\nonumber\\
&=&1+\frac{1}{2^{N-1}}\biggr[\sqrt{1+(V\sin\xi)^2}\sin(\theta_{11}+\Delta)-1\biggr],\nonumber\\
\end{eqnarray}
with $\Delta=-\arctan\frac{1}{V\sin\xi}$. When
$2\arctan\frac{1}{V\sin\xi}<\theta_{11}<\pi$, one has $\mathcal
{I}^N_{\rm QM}>1$. Therefore the violation of Bell's theorem always
exists for the odd-$N$ case. This ends the proof.


In summary, starting from an input single-qubit Schr{\"o}dinger's
cat state, the non-existence of perfect $1\rightarrow N$ QCM can
always be certified by Bell's theorem even in the presence of noise.
Because of the no-cloning theorem, the input one-qubit
Schr{\"o}dinger's cat state cannot be perfectly copied, and this
results in the violation of Bell's inequalities by the output state.
It is noted that the relation between Bell's inequality and quantum
no-cloning theorem has also been investigated in a different way
\cite{W01, nosignal}. Starting from the assumption that Alice and
Bob share nonlocal correlations, Werner showed that the violation of
CHSH inequality leads to quantum no-cloning theorem \cite{W01}.
Later Masanes \emph{et al}. extended the result to nonsignaling
theories \cite{nosignal}.
Different from the above literatures, our investigation begins from
the single-qubit Schr{\"o}dinger's cat state that does not have the
assumption of original correlation between subsystems. Our result
demonstrates that the no-cloning theorem is fundamentally linked to
quantum nonlocality and gives new sights into the nature of quantum
no-cloning theorem. We anticipate our work will initiate further
exploration of no-cloning theorem and quantum nonlocality for the
purpose of better understanding our quantum physical world.

J.L.C. is supported by
National Basic Research Program (973 Program) of China under Grant
No. 2012CB921900 and NSF of China (Grant Nos. 10975075 and
11175089). This work is also partly supported by National Research
Foundation and Ministry of Education, Singapore (Grant No. WBS:
R-710-000-008-271).



%

\end{document}